\documentclass[twocolumn,prl,aps,showpacs,floatfix,superscriptaddress]{revtex4}
\usepackage{epsfig}
\usepackage{graphicx}
\begin{document}

\noindent {\bf Comment on ``Influence of Noise on Force Measurement"}

\vspace{0.3cm}

In a recent Letter \cite{Volpe:10}, Volpe et al.\ describe experiments on a
colloidal particle near a wall in the presence of a gravitational field for
which they study the influence of noise on the measurement of force. Their
central result is a striking discrepancy between the forces derived from
experimental drift measurements via their Eq.\ (1), and from the equilibrium
distribution. From this discrepancy they infer the stochastic calculus realised
in the system.

We comment, however: (a) that Eq.\ (1) does not hold for space-dependent
diffusion, and corrections should be introduced; and (b) that the ``force"
derived from the drift need not coincide with the ``force" obtained from the
equilibrium distribution.

The problem of what should be the ``correct" stochastic calculus was tackled
in the early 1980s. The consensus was that, for a model in the form of a
stochastic differential equation (SDE), the calculus to be used, e.g.\ in a
simulation, is part of the model itself. Correspondingly, starting from
measured data, what we observe is a distribution function;
but in the absence of further information and/or specific models, we cannot
infer the underlying stochastic calculus \cite{VanKampen:81}. For a continuous
physical system, with noise of (inevitably) finite bandwidth, we expect the
Stratonovich calculus to apply \cite{Smythe:83}.

From the SDE Eq.\ (3) of~\cite{Volpe:10}
\begin{equation}
dz = f(z) dt + g(z) dW  = \frac{F(z)}{\gamma(z)} dt + \sqrt{2 D_\perp (z)} dW
\label{sde}
\end{equation}
we obtain the family of Fokker-Plank equations
\begin{equation}
\frac{\partial P(z,t)}{\partial t} = \frac{\partial}{\partial z}\left\{-f(z)
-\alpha g(z) g'(z)+ \frac{1}{2} \frac{\partial}{\partial z} g^2(z)\right\} P(z,t)
\label{fp}
\end{equation}
where $g'(z)=\partial g(z)/\partial z$, and $\alpha$ is 0 or 1/2 for the Ito or
Stratonovich stochastic calculi respectively. In an experiment the diffusion
(related to $g^2(z)$) and the drift (related to $f(z) + \alpha g(z) g'(z)$) can
be measured. To infer $\alpha$, however, additional information (e.g.\
knowledge of $f(z)$) is needed. From (\ref{fp}), the drift velocity is
\begin{equation}
\overline{v}_d = \frac{dz}{dt} = f(z) + \alpha g(z) g'(z) = \frac{F(z)}{\gamma(z)} + \alpha \frac{d D_\perp(z)}{d z}
\label{f1}
\end{equation}
This relation does not coincide with Eq.\ (1) of~\cite{Volpe:10} because the
nonlinearity of the diffusion coefficient enters the drift. Hence, it is
impossible to derive the ``force" $F(z)$ from a measurement of the drift
velocity, as in Eq.\ (1) of~\cite{Volpe:10}, where it was assumed that $F(z) =
\gamma(z) \overline{v}_d(z)$.

\begin{figure}
\includegraphics*[width=6.5 cm]{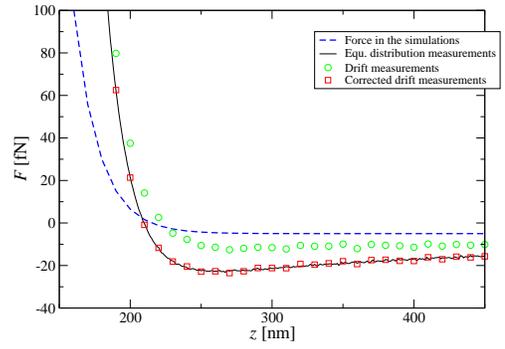}
\caption{(color online) Force computed from a simulation of (\ref{sde}) using
the Stratonovich calculus. Consistent with the experiment, we took $f(z) =
Be^{-kz} + C$ with $B=770\;\rm{pN}$, $C=-5\;\rm{fN}$, $k=(18 \;\rm{nm})^{-1}$,
and $D_\perp(z) = D_\infty z/(z+a)$ with $z$ in nm, $a=700\;\rm{nm}$, $D_\infty=k_B
T/6 \pi \eta R$, $2R=1.31 \;\rm{nm}$, $T=300\; \rm{K}$, $\eta = 8.5\times
10^{-3} \;\rm{Pa\;s}$.} \label{fig1}
\end{figure}

In~\cite{Volpe:10} the force is also computed from $F_e(z) = -d U(z)/dz$ where
the potential $U(z) = -k_B T \ln(P(z))$ is obtained from the the equilibrium
distribution $P(z)$. From~(\ref{fp}) we obtain
$$U(z) = - k_B T \int \frac{f(z) + (\alpha-1) g(z)g'(z)}{g^2(z)/2}\,dz$$
\begin{equation}
\frac{F_e(z)}{\gamma(z)}= -\frac{1}{\gamma(z)}\frac{d U(z)}{dz} = f(z) + (\alpha-1) g(z)g'(z)
\label{f2}
\end{equation}
Eqs.~(\ref{f1}) and~(\ref{f2}) differ by $-g(z)g'(z)=-\frac{d D_\perp(z)}{d
z}$, which is \textit{exactly} the experimental discrepancy reported
in~\cite{Volpe:10}; this difference is independent of $\alpha$, i.e.\
independent of the stochastic calculus used to describe the physical system.

As a demonstration, we simulated (\ref{sde}) numerically for the Stratonovich
calculus with the same definitions as~\cite{Volpe:10}, computing the average
forces from the drift and equilibrium distribution of the time sequence $z(t)$.
Fig.~\ref{fig1} shows that the force from the drift ($\bar v_d \gamma(z)$,
Eq.~(\ref{f1}), circles) differs from that from the equilibrium distribution
($F_e(z)$, Eq.~(\ref{f2}), full curve). When the drift result is corrected by
the additional term $-g(z)g'(z)$, however, we recover the equilibrium
distribution result. Thus the discrepancy reported in~\cite{Volpe:10} has
nothing to do with different stochastic calculi: it is simply a consequence of
having two different definitions of ``force''. Neither of them corresponds to
the true microscopic force, and they coincide only where the diffusion
coefficient happens to be constant.

We are grateful to Ping Ao for alerting us to this problem and to Mark Dykman
for valuable discussions.

\vspace{0.2cm}

\noindent R.~Mannella,$^1$ P.~V.~E.~McClintock,$^2$

\noindent$^1$Dipartimento di Fisica, Universit\`a di Pisa, \\I-56127 Pisa, Italy\\
$^2$Department of Physics, Lancaster University, \\Lancaster LA1 4YB, UK

\vspace{0.2cm}



\end{document}